\documentclass[english]{article}
\usepackage[utf8]{inputenc}
\usepackage[T1]{fontenc}
\usepackage{babel}
\usepackage{amsmath}
\usepackage{graphicx}
\usepackage{fancyhdr}

\usepackage{setspace}

\usepackage[dvipsnames,table]{xcolor}
\usepackage[colorlinks]{hyperref}
\usepackage{multirow}
\usepackage{threeparttable}
\usepackage[margin=1.5in]{geometry}

\usepackage[utf8]{inputenc}
\usepackage{amsmath}
\usepackage{framed}
\usepackage[shortlabels]{enumitem}
\usepackage{graphicx, caption}
\usepackage{float}
\usepackage{textcomp}
\usepackage{gensymb}
\usepackage{mathtools}
\usepackage{rotating}
\usepackage{subcaption}
\usepackage{bm}
\usepackage{amssymb}
\usepackage{cancel}
\usepackage{changes}

\usepackage{listings}
\usepackage{color}

\definecolor{dkgreen}{rgb}{0,0.6,0}
\definecolor{gray}{rgb}{0.5,0.5,0.5}
\definecolor{mauve}{rgb}{0.58,0,0.82}

\DeclareUnicodeCharacter{2212}{-}

\title{Text-mined dataset of gold nanoparticle synthesis procedures, morphologies, and size entities}
\author{Kevin Cruse\textsuperscript{1,2}, Amalie Trewartha\textsuperscript{2}, Sanghoon Lee\textsuperscript{1,3}, \\
Zheren Wang\textsuperscript{1,2}, Haoyan Huo\textsuperscript{1,2}, Tanjin He\textsuperscript{1,2}, Olga Kononova\textsuperscript{1,2}, \\
Anubhav Jain\textsuperscript{3}, Gerbrand Ceder\textsuperscript{1,2{*}}}

\begin{document}
\date{}
\maketitle

\noindent\textsuperscript{1}Department of Materials Science and Engineering, University of California, Berkeley, CA 94720, USA; \\
\textsuperscript{2}Materials Sciences Division, Lawrence Berkeley National Laboratory, Berkeley, CA 94720, USA; \\
\textsuperscript{3}Energy Technologies Area, Lawrence Berkeley National Laboratory, Berkeley, CA 94720, USA; \\
{*}Corresponding author: Gerbrand Ceder (gceder@berkeley.edu)\\

\newpage

\section*{Abstract}

Gold nanoparticles are highly desired for a range of technological applications due to their tunable properties, which are dictated by the size and shape of the constituent particles. 
Many heuristic methods for controlling the morphological characteristics of gold nanoparticles are well known. However, the underlying mechanisms controlling their size and shape remain poorly understood, partly due to the immense range of possible combinations of synthesis parameters.  
Data-driven methods can offer insight to help guide understanding of these underlying mechanisms, so long as sufficient synthesis data are available. To facilitate data mining in this direction, we have constructed and made publicly available a dataset of codified gold nanoparticle synthesis protocols and outcomes extracted directly from the nanoparticle materials science literature using natural language processing and text-mining techniques. This dataset contains 5,154 data records, each representing a single gold nanoparticle synthesis article, filtered from a database of 4,973,165 publications. Each record contains codified synthesis protocols and extracted morphological information from a total of 7,608 experimental and 12,519 characterization paragraphs.

\newpage
 
\section*{Background \& Summary}

The synthesis of gold nanoparticles has been practiced for centuries, and their modern applications are widespread, which include in vitro diagnostics\cite{CancerDetect}, semiconductor technology\cite{Semicond}, and cosmetics\cite{Cosmetics}. The application of gold nanoparticles often depends on their morphology and size\cite{PVPRod}; yet, despite their ubiquity, only relatively recently has the control of these properties been interrogated systematically\cite{AuNPSynthReview}. 

While many theories and models exist for the mechanisms that determine nanoparticle morphology \cite{AuNPShapeControl,AuNPShapeControl2,FormationModel}, most of the exploration of this synthesis space is driven by heuristics. For nanorod growth in particular, it appears that the simultaneous presence of many reagents affects the final characteristics of a sample of gold nanorods\cite{AuNRReview2}. While factorial experiments can offer some insights into how varying certain precursor concentrations affects final particle morphology, size, or aspect ratio, it is impractical to perform enough experiments to cover a large enough portion of the synthesis space to produce an effective model, even with state-of-the-art high-throughput synthesis methods. 

Beyond empirical modeling and experiment, computational methods exist that either simulate the energetics of the formation of nanoparticles or interrogate the nucleation and growth steps traversed by nanoparticles. However, these approaches come with inherent tradeoffs between the resolution of atomic interaction and computational tractability. 
For example, calculations from first-principles have been conducted using density functional theory (DFT) that probe the energetic landscape of potential gold nanoparticle shapes\cite{DFTAu12}, including the effects of various surface ligands\cite{DFTLigand}, which are vital for the synthesis of solution-phase noble metal nanoparticles\cite{ProtMole}. However, such a technique does not take into account the intricacies of nucleation and growth competition in solution-based nanoparticle synthesis. 
On the other hand, continuum-level models can represent real-time growth and dispersity dynamics\cite{Continuum}, though sacrificing the small-scale energetics highlighted by techniques such as DFT. 

In a third paradigm of scientific investigation, the volume of data-driven approaches to understanding chemistry and materials synthesis is accelerating. These approaches represent a resourceful complement to established computational methods and raw experimentation, and have been proven successful in applications such as materials discovery \cite{MatDiscovery, FisherCrystalStructure}, synthesis protocol querying \cite{WestonNER}, and the simulation and interpretation of characterization results \cite{AutoDetect, SzymanskiXRD}. 
However, data-driven approaches are limited by the completeness and substance of the data resource(s) used. For the nanotechnology field, Yan et al. \cite{NPStructureDb} constructed a table of available nanotechnology databases, their volume, as well as a description of their proposed usages. Each of these, including the dataset created by Yan et al., provide useful and carefully curated information on the characteristic features of various nanostructures; however, to our knowledge, none provide validated protocols for their synthesis. 

There is a wealth of gold nanoparticle synthesis and characterization data available for data-driven approaches in unstructured form in the scientific literature \cite{LLNLNano}. This source remains largely untapped because manually extracting such information is both tedious and unrealistic due to the volume of the literature. Natural language processing (NLP) and text-mining techniques have been successfully employed and established in scientific fields related to materials science \cite{KrallingerIR2017, ChemTagger, CDE}, and the sub-field of text-mining in materials science is budding \cite{KononovaTMReview, OlivettiReview, KimInsights, Tshitoyan2019}. Here we employ natural language processing (NLP) and text-mining techniques on a collection of nearly 5 million materials science publications \cite{KononovaSolidState} to extract gold nanoparticle synthesis recipes and their outcomes. We introduce this dataset as an additional element, along with direct experimentation and computational modeling, in the effort to capture the mechanisms for metal nanoparticle growth.

\begin{figure}[ht]
\centering
\includegraphics[width=\linewidth]{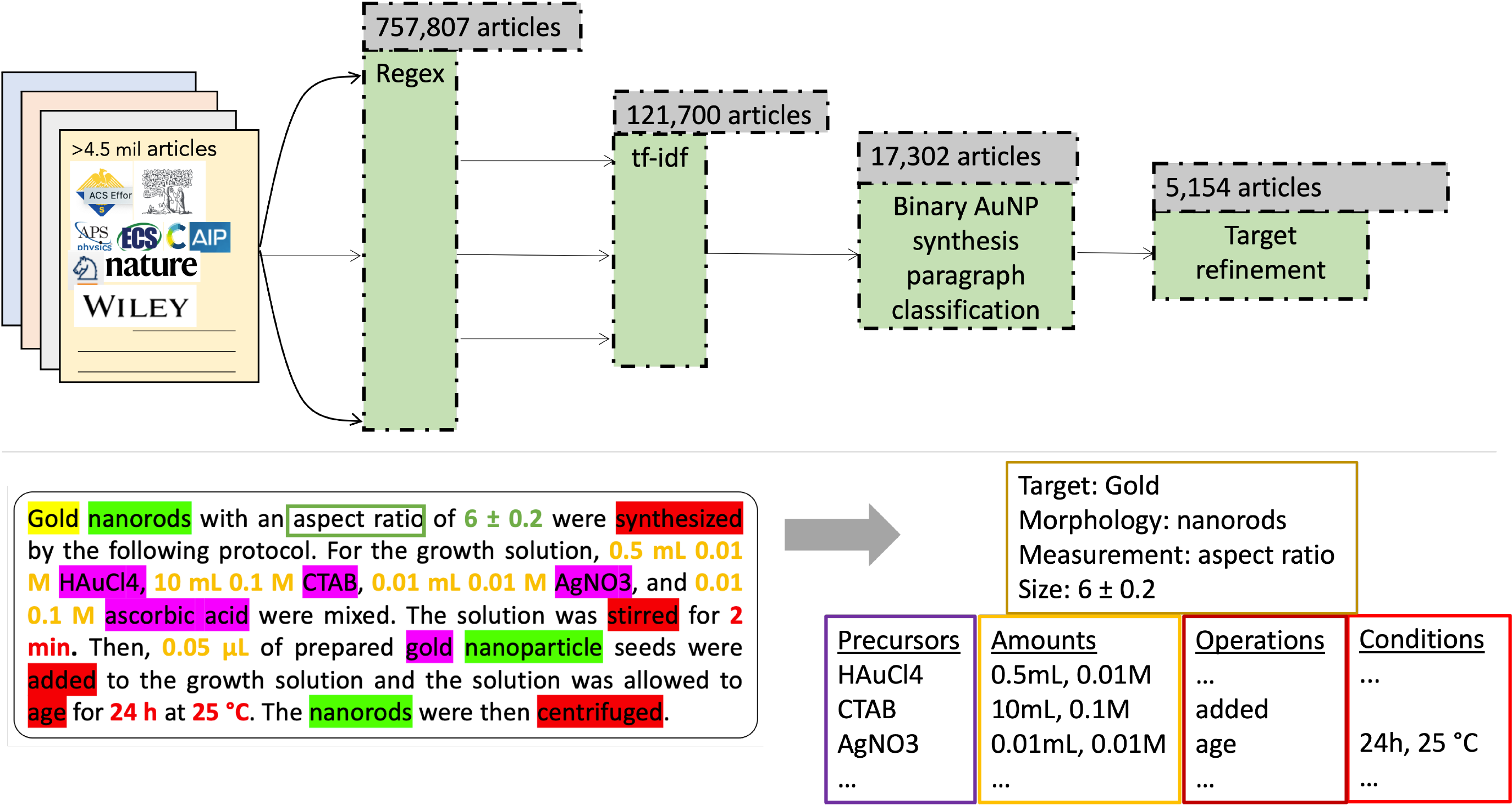}
\caption{
\textbf{AuNP synthesis publication extraction pipeline.} Starting with the >4.5 million article materials science literature database, parsed paragraphs from the articles are funneled through progressively finer-meshed filters to identify those related to the synthesis of gold nanoparticles. The first two steps include a regex search for "nano\underline{\hspace{1cm}}" phrases followed by the vectorization of that corpus using tf-idf, similar to the method used in Hiszpanski et al. \cite{LLNLNano}. Those articles where the tf-idf scores for gold are higher than other noble metal nanoparticle compositions are accepted. Each paragraph from those articles is then passed through a binary classifier which determines whether or not the paragraph describes the synthesis of gold nanoparticles. Finally, after extracting the synthesis recipes from those relevant paragraphs, 5,154 articles with synthesis paragraphs containing gold- or gold nanoparticle-related targets are collected. An example synthesis paragraph along with a sample of the extracted information is shown in the bottom panel. 
}
\label{fig:aunp_pipe}
\end{figure}

In this data descriptor, we present an open-source dataset of 5,154 records filtered down from a database of 4,973,165 publications, each representing a single article and containing automatically extracted gold nanoparticle synthesis recipes and morphological information from the contained synthesis paragraphs and characterization paragraphs. Within a synthesis paragraph, the precursors used and their amounts are extracted, and the synthesis actions and conditions are extracted and codified into a procedure graph. Within both synthesis paragraphs and characterization paragraphs, morphological (e.g. ``spherical'', ``nanorod'', ``dogbone-like'') and size (e.g. diameter, aspect ratio) entities were extracted. In total, this dataset encompasses 7,608 synthesis paragraphs and 12,519 characterization paragraphs. A schematic of the pipeline devised for this extraction is shown in Figure \ref{fig:aunp_pipe}. 

\section*{Methods}

\subsection*{Content Acquisition}
\label{section: cont_acq}

Over 4.5 million publications from material science journals have been scraped from the web and parsed by the process described in Kononova et al. \cite{KononovaSolidState}. These publications were obtained through agreements with publishers Elsevier, Wiley, the Royal Society of Chemistry, Nature Publishing Group, the American Institute of Physics, Springer, the American Chemical Society, the American Physical Society, and the Electrochemical Society. Journals from each publisher related to materials science were selected manually. Articles were scraped from these publishers using a custom web scraping tool. Only articles published after the year 2000 were collected because prior publications are often provided in PDF and not HTML/XML format, the latter of which is more straightforward to parse \cite{KononovaTMReview}. Parsers were built for each publishing group using custom parsing toolkits. From each article, the full text and relevant metadata were parsed and are stored across MongoDB database collections (\url{www.mongodb.com}) maintained on an internal server.

\subsection*{AuNP Synthesis Publication Collection}
\label{section: pub_col}

Publications containing gold nanoparticle synthesis procedures were collected using a series of progressively finer-meshed filters, combining unsupervised and supervised text processing methods. This progression is illustrated in Figure~\ref{fig:aunp_pipe}, along with the corresponding yield of articles at each step. The stages of the pipeline were built to facilitate adaptation to other nanomaterial compositions in the future. Each of the steps is described in detail below. 

\subsubsection*{Identifying AuNP Publications}
\label{section: pub_filt}

The early stages of the pipeline cast a wide mesh for Au nanoparticle (AuNP) publications using unsupervised text-mining methods. These first two steps were adapted from the methods used for nanomaterial synthesis publication extraction performed by Hiszpanski et. al \cite{LLNLNano}. The first step consists of a simple regular expression query.
To accomplish this, we imported the paragraphs of all papers from our MongoDB database (described in ``Content Acquisition'') into an Apache Solr search engine instance (\url{https://solr.apache.org/}), which is powered by Apache Lucene and supports fast full-text search. We used the default English Lucene analyzers and tokenizers in Solr. The full text of every article was queried for words starting with ``nano'', followed by any number of characters or whitespace. The query was written in Solr-specific query syntax and returned identifiers of the matching paragraphs. We used the query results to lookup and fetch related papers in the MongoDB database. This step yielded 757,807 nanomaterial articles.

Next, all of the documents in this corpus were vectorized in a term frequency-inverse document frequency (TF-IDF) representation, where each element in the vector represents the frequency of a given token normalized by the frequency of that term across documents \cite{jurafskynlp}. We implemented this vectorization using the scikit-learn \texttt{TfidfVectorizer} module (\url{https://scikit-learn.org/stable/}), with tokenization from ChemDataExtractor's \texttt{ChemWordTokenizer} \cite{CDE}, default English stopwords eliminated, and a minimum document frequency of 100 for each token. With the TF-IDF vectorized corpus, articles were collected whose TF-IDF values for the words ``gold'' or ``Au'' were larger than any of ``silver'', ``Ag'', ``copper'', ``Cu'', ``palladium'', ``Pd'', ``platinum'', or ``Pt''. 121,700 gold nanomaterial articles were collected in this step, each being related to gold nanomaterials but not necessarily containing protocols for their synthesis.

\subsubsection*{AuNP Synthesis Paragraph Classification}
\label{section: synth_class}

To isolate those AuNP publications that contain synthesis protocols, we trained a transformer-based binary classification model using the Simple Transformers NLP library (\url{https://github.com/ThilinaRajapakse/simpletransformers}). We first pre-trained a BERT \cite{BERT} (Bidirectional Encoder Representations from Transformers) model specializing in materials science text, referred to as ``MatBERT'' \cite{MatBERTNER}. 
The pre-training data for MatBERT were 2 million randomly sampled papers from our publications database. Following the original BERT, we trained two WordPiece tokenizers of vocabulary size 30,522 (cased and uncased) from scratch on this full-text to optimize tokenization for materials science terminologies. After all the papers were tokenized, paragraphs with less than 20 or more than 510 tokens were removed. Out of 61 million paragraphs from the 2 million sampled papers, roughly 17\% contained less than 20 tokens and about 2\% contained more than 510 tokens, for both the cased and uncased set of texts. This yielded around 50 million paragraphs and 8.8 billion tokens. During pre-training, MatBERT was trained for the masked language modeling (MLM) task, which requires MatBERT to predict the original tokens in a paragraph after they are masked. This pre-training step helps MatBERT to develop a general understanding of the language and better learn the classification of synthesis protocols. The training codes and pre-trained MatBERT models can be found at \url{https://github.com/lbnlp/MatBERT}.

To gather positive training paragraphs, we first modeled the topics of every paragraph in the aforementioned gold nanoparticle publication collection using latent Dirichlet allocation (LDA) \cite{LDA}. Then, we collected and manually validated those paragraphs whose dominant topic was related to synthesis (topic words including ``synthesized'', ``solution'', ``ml'', ``addition'', etc.). A range of negative training paragraphs were collected manually from various parts of a typical publication, including the introduction, results, discussion, and characterization sections. Annotations were accomplished using SpaCy's Prodigy interface (\url{https://prodi.gy}). The training data ultimately included 739 training examples, with 242 positive examples and 497 negative examples. Because synthesis paragraphs are far less common in literature than non-synthesis paragraphs, we included more negative than positive training examples to ensure that most kinds of these non-synthesis paragraphs were covered in the training data. Using the \texttt{ClassificationModel} module from Simple Transformers, training data was split into 80/10/10 train/validation/test sets and trained over 20 epochs. Articles were then identified that contained at least one paragraph classified as being related to gold nanoparticle synthesis. This step yielded 21,989 AuNP synthesis paragraphs from a total of 17,302 articles. 

\subsubsection*{Synthesis Recipe Extraction}
\label{section: rec_ext}

Synthesis targets, precursors, their amounts, synthesis actions, and action conditions were all extracted using synthesis procedure extraction and codification tools described in \cite{KononovaSolidState} and \cite{HeCER}. A sample of an example extraction from a synthesis paragraph is shown in the bottom panel of Figure \ref{fig:aunp_pipe}. Each step is described in detail below.

\emph{Materials entity recognition (MER).} 
To identify and classify targets, precursors, and other materials from synthesis paragraphs, we implemented a two-step model.
In the first step, each word token was transformed into an embedding vector with the MatBERT model (see ``AuNP Synthesis Paragraph Classification'').
Then, the embedding vector was passed to a bi-directional long-short-term memory neural network with a conditional random-field top layer (BiLSTM-CRF) to identify whether the corresponding token was a materials entity or a regular word. 
In the second step, each materials entity was replaced with a keyword \emph{<MAT>} and classified as either a \emph{target}, \emph{precursor}, or \emph{other} material using another BERT-based BiLSTM-CRF network with a similar structure.
In total 1,281 synthesis paragraphs from 1,155 papers were annotated by labeling each word token as \emph{material}, \emph{target}, \emph{precursor}, or \emph{outside}. 
The annotated dataset was split into training/validation/test sets with a paper-wise ratio of 700/150/305 to train the aforementioned two neural networks.

\emph{Synthesis actions and their attributes.}
To recognize and classify synthesis actions described in a paragraph, we implemented an algorithm that combines a recurrent neural network (RNN) and rule-based parsing of sentence dependency trees. Sentences were tokenized using ChemDataExtractor's \texttt{ChemWordTokenizer}. The RNN performed classification of sentence tokens into 5 categories: \emph{start-synthesis} (general actions that signify that something was synthesized, e.g. ``synthesized'', ``prepared'', etc.), \emph{mixing}, \emph{heating}, \emph{drying}, and \emph{cooling}, which are the basic actions in nanoparticle synthesis. The RNN was trained on a set of 3,040 synthesis sentences from 535 synthesis paragraphs (classified according to the paragraph classifier described in Huo et al. \cite{HuoParaClass}). In brief, 3,781 sentences were taken from 199 solid-state, 51 sol-gel, 148 hydrothermal, and 137 precipitation synthesis paragraphs. 3,040 sentences in this set were determined to be synthesis sentences (as opposed to characterization or miscellany). The tokens in these 3,040 synthesis sentences were annotated by human experts in NLP and materials synthesis science according to their type of synthesis action, with the actions relevant to nanoparticle synthesis listed above.
The tokens' feature vectors were generated using a Word2Vec model \cite{mikolov2013distributed}.
The embeddings were trained on $\sim$400,000 synthesis paragraphs of different synthesis types using the Gensim library \cite{rehurek_lrec}.
The sentences of paragraphs were lemmatized, all the quantity tokens were replaced with the keyword <NUM>, and all the chemical formulas were replaced with the keyword <CHEM> using rule-based algorithms.
The SpaCy library \cite{SpaCy} was used to grammatically parse each sentence and obtain linguistic features of the tokens, such as their part of speech and their dependency on root tokens.
For training, validation and testing, the annotated set was split into a 70/10/20 fraction, respectively.
Synthesis action attributes, such as temperature, time, and environment were extracted by using dependency tree parsing and a rule-based regular expression approach \cite{jurafskynlp}. 


\emph{Material quantities extraction.}
To correlate the numerical values of material quantities, such as molarity, concentration, or volume, to materials entities extracted by the MER model (see above), we applied a rule-based approach.
First, we used the NLTK library \cite{NLTK} to build syntax trees \cite{jurafskynlp} for each sentence in a paragraph, where every word is represented as a leaf node.
Then, the syntax tree for each sentence was cut into the largest sub-trees for every material, with each sub-tree having only one material entity. To do this, we first identified the materials on leaf nodes. Then, starting from each material, we identified the largest sub-trees (i.e., we traversed the syntax tree upwards until there was more than one material leaf node descending from the same node). Finally, the largest sub-tree for a given material was defined as the sub-tree formed by the node and its descendants identified in the previous step.
Next, we searched for the quantities in each sub-tree and assigned the quantities associated with the unique material entity in the sub-tree.

\subsubsection*{Gold-related Target Refinement}

With all relevant information from synthesis paragraphs extracted and codified, we implemented a final target refinement step to identify all of those papers that contain synthesis procedures explicitly targeting ``gold'' or gold nanoparticle-related entities, a list for which is provided in the \texttt{/rsc} folder of the GitHub codebase (see ``Code Availability''). Of the 18,101 articles collected from the binary classification step described above, 5,154 contained a target entity extracted by MER (see ``Synthesis Recipe Extraction'') that was related to gold or a gold nanoparticle-related entity. 

\subsubsection*{Tagging Seed-mediated Growths}

Each paragraph containing a recipe was tagged as being related to either a seed-mediated or seedless synthesis approach. Seed-mediated approaches are those in which some method is used to create small colloidal seeds that act as nucleation sites for larger growths, often with interesting morphology. These methods are common for rod-based nanoparticle growths and are abundant in the literature \cite{AuNRReview}, so we wanted to make those particular recipes easily queryable. The tags for this field were determined by keyword matching for ``seed'' and related lemmas for seed-mediated methods as well as ``seedless'' or the absence of seed-related text for seedless methods. The binary tag for seed-mediated growth in a given recipe paragraph is included in the \texttt{seed\_mediated} field as a boolean in the provided dataset (see Table \ref{tab:tab3}).

\subsection*{AuNP Synthesis Outcome Extraction}

To complement the extracted gold nanoparticle synthesis protocols from the collected 5,154 articles, we also extracted relevant morphological and characterization information. A two-step process was implemented for this extraction. 

\subsubsection*{Characterization Paragraph Classification}
\label{section: char_class}

To focus on paragraphs that contain information on the morphology of synthesized gold nanoparticles, we trained a binary transformer-based gold nanoparticle characterization paragraph classifier, using a similar approach to the binary synthesis paragraph classifier described above. Positive training paragraphs were collected by manually selecting characterization-related and morphology-related paragraphs modeled from LDA (with topic words including ``morphology'', ``tem'', ``size'', ``diameter'', ``nm'', etc.). The training data for this classifier included 299 training examples, with 69 positive examples and 230 negative examples. Using an 80/10/10 train/validation/test split, a model was trained over the course of 20 epochs using Simple Transformer's \texttt{ClassificationModel} Module. This classification yielded 12,519 paragraphs containing morphological characterization information for gold nanoparticles. 

\subsubsection*{Characterization Entity Recognition (MorphER)}
\label{section: char_ext}

To extract relevant gold nanoparticle morphological information, we developed a transformer-based named entity recognition (NER) model specializing in the recognition of entities related to nanoparticle morphology and size (``MorphER''). To train this model, we annotated a set of 119 characterization-classified paragraphs from 91 articles on gold nanoparticle synthesis. The entities labeled for this model include specific morphological information for the synthesized gold nanoparticles, including: \textbf{MOR}, noun phrases related to morphology, such as ``nanoparticles'' or ``AuNRs''; \textbf{DES}, descriptive terms for morphologies, such as ``dumbbell-like'' or ``spherical''; \textbf{MES}, measurements, such as ``aspect ratio'' or ``diameter''; \textbf{SIZ}, the value of the measurement; \textbf{UNT}, unit, if applicable (i.e. not for aspect ratios).
Entities related to nanoparticles (e.g. ``NP'', ``anoparticle'') but not necessarily their shape were labeled as \textbf{MOR} entities since the shape of the particle is not always mentioned explicitly, though the size is usually mentioned. This way, one could attribute extracted size information to at least some target entity. 
We chose to use NER to extract size information as well to deal with cases where we cannot use units as an anchor for rule-based methods, as in aspect ratios for nanorods. The model was fined-tuned over the pretrained MatBERT model described earlier (see ``AuNP Synthesis Paragraph Classification'') on the paragraph-level with an 80/10/10 train/validation/test split over 20 epochs and deep fine-tuning. This entity recognition model was run on any paragraph that the AuNP synthesis paragraph classifier predicted to be a synthesis paragraph or that the characterization paragraph classifier predicted to be a characterization paragraph. 

This is the final extraction step in the pipeline constructed to build this dataset. Thus, the dataset does not contain any entity linking (e.g. particle size to specific nanoparticle morphology, morphological entity to synthesis procedure, etc.). In attempts to address this for the next iteration of the dataset, we have implemented, with moderate success, both rule-based linking through dependency tree parsing as well as simultaneous extraction and linking using more powerful language models such as GPT-3.

Briefly, we address our decisions for and differences in model choice and architecture for the text entity extraction tools described above. Materials Entity Recognition \cite{HeCER} and the synthesis actions extraction model were first trained for the extraction of inorganic solid-state synthesis procedures \cite{KononovaSolidState}. The development of our MatBERT model (described in ``AuNP Synthesis Paragraph Classification'') was more recent and coincided with the development of the Characterization Entity Recognition model. Since our development of MatBERT, we have incorporated its embeddings into the Materials Entity Recognition model since this tool is used on paragraphs outside of synthesis paragraphs. Because the extraction task for synthesis actions is linguistically simpler than for materials’ names, we continue to use the Word2Vec embeddings trained for synthesis action extraction. Using MatBERT is also significantly more time consuming (as determined by He et al. \cite{HeCER}) and Word2Vec embeddings are sufficient for modeling word similarity. Additionally, the RNN model used for synthesis action extraction is capable of capturing contextual differences for certain vocabulary.

\section*{Data Records}
The dataset, with 7,608 synthesis paragraphs and 12,519 characterization paragraphs from 5,154 articles, is provided as a JSON file, available publicly at \url{https://doi.org/10.6084/m9.figshare.16614262.v3} \cite{AuNP_Dataset}. Each record corresponds to a publication, represented as a JSON object in a top-level list. Within each record is a list of paragraphs, with some containing a codified recipe, extracted morphological information, both, or neither. Metadata contained in the dataset for an article include: article DOI, the year of publication, and the number of times the article has been cited as of August 2021. For each paragraph within an article, metadata include: a unique paragraph hash, a boolean indicating whether or not the paragraph contains synthesis, a boolean indicating whether or not the paragraph contains characterization information, a boolean indicating whether or not the paragraph contains a seed-mediated growth, and a snippet of the paragraph text. Expanded details for the format of the dataset are given in Tables \ref{tab:tab2},  \ref{tab:tab3}, and \ref{tab:tab4}.

\section*{Technical Validation}
\label{section: tech_val}

The quality and content of this dataset is evaluated below through a description of the data extraction model metrics as well as a comparison of the dataset demographics to established heuristics in the field. 

\subsection*{Extraction Accuracy}

We use the 35k article nanomaterial dataset developed by Hiszpanski et al. \cite{LLNLNano} as a benchmark with which to compare our regex/tf-idf article filtering. We note that this dataset is comprised only of publications from Elsevier, so we only evaluate model performance on this set, which comprises 69\% of our total materials science literature collection. The 35k article gold standard nanomaterial dataset contains 10,229 articles predominantly related to gold, of which 2,577 are not contained in our original MongoDB collection and 602 were not captured by our tf-idf method. Inspection showed that the volume of articles not contained in our database is largely due to our journal selection during the scraping and parsing of articles, which focuses on materials science-specific journals (whereas  Hiszpanski et al. selected from all publications in Elsevier journals). For the 602 articles not captured by our data filtering, it was found on manual inspection that many only mentioned ``nano\underline{\hspace{1cm}}'' or ``gold''-related vocabulary once or twice throughout the article or only in the abstract. Such articles are not considered valuable for this dataset since they likely do not contain recipes for gold nanoparticle synthesis, so their absence is appropriate. No false positives (i.e. articles that our pipeline determined to be related to gold nanomaterials but that Hiszpanski et al. determined to be related to another composition) were found from our extraction.

Manual validation was previously performed for 100 solution-based synthesis paragraphs for another recently accepted dataset manuscript \cite{SolutionDataset}. This was done to determine the extraction accuracy of the rule-based methods used in the extraction pipeline, which included synthesis action conditions (time and temperature) as well as the amounts of materials used. These metrics are included for reference in Table \ref{tab:tab5} as well. We accepted scores with higher precision than recall for these rule-based methods in order to avoid contaminating the dataset with incorrect information, though potentially sacrificing completeness of a given codified recipe. 

Manual checks on the validity of the seed-mediated growth tag for 50 paragraphs were performed, including 25 on paragraphs determined to contain a seed-mediated growth method and 25 on paragraphs containing seedless growth. 49 out of the 50 checks were determined to be valid and true. 1 paragraph was labeled as ``seedless'', though it only contained purchasing information. We still considered this tagging valid since the incorrect classification is due to the synthesis paragraph classifier earlier in the pipeline. The accuracy for this tagging method is shown in \ref{tab:tab5}. 

Finally, the F1 score, precision, and recall for each of the paragraph classifiers and the MorphER model (along with the F1 score, precision, and recall for each of the constituent entities) are also shown in Table \ref{tab:tab5}. For the binary classification models, similarly to the rule-based methods discussed above, we accepted scores with higher precision than recall in order to avoid erroneous classifications of paragraphs that should be data-rich, and thus avoiding inflating the breadth of the present dataset.

\subsection*{Dataset Mining}
\label{section: datamining}

The statistical breakdown of the recipes contained in this dataset are visualized and some basic correlations are explored. These are then discussed in the context of current knowledge in the field. 

\begin{figure}[ht]
\centering
\includegraphics[width=\linewidth]{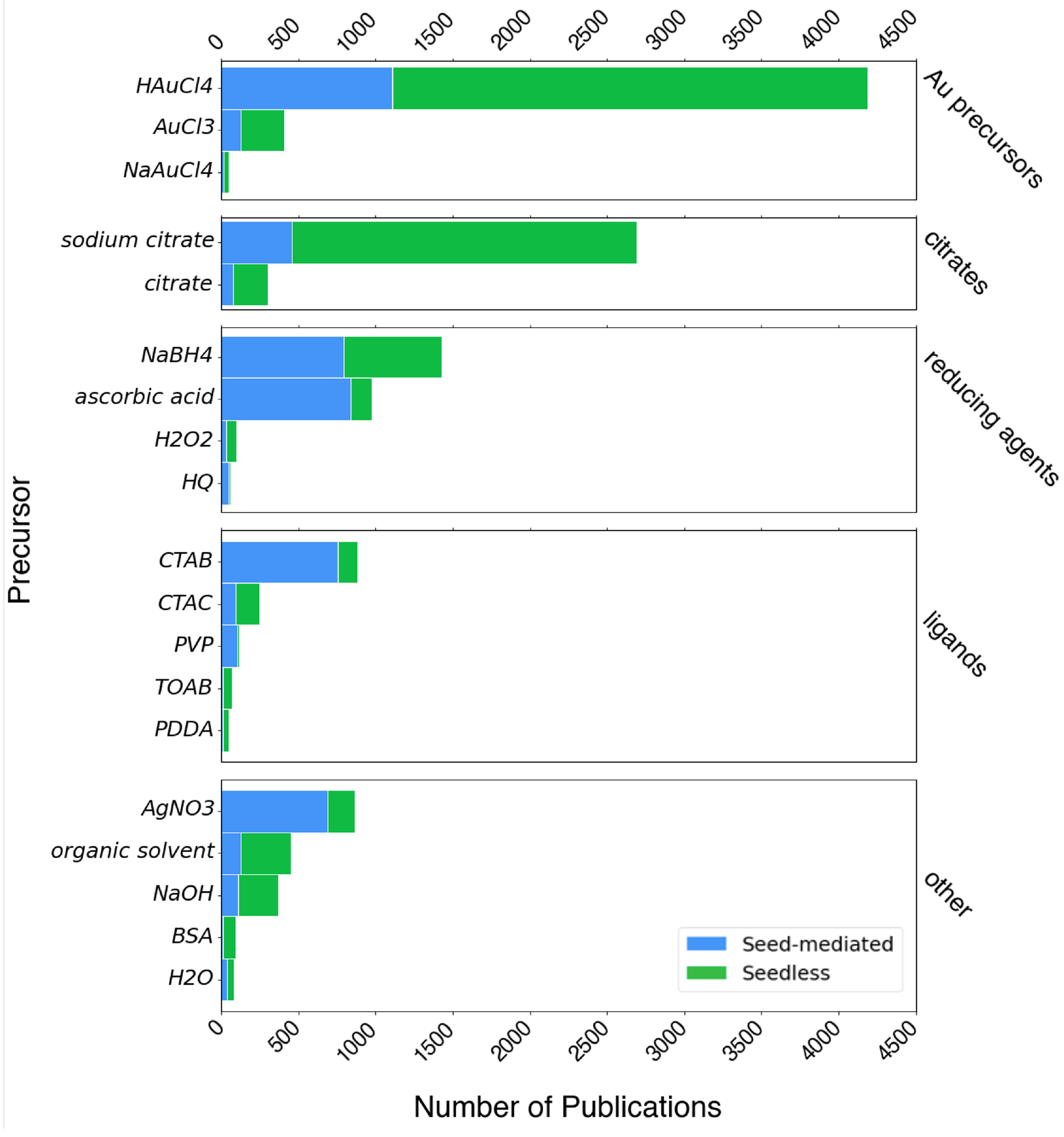}
\caption{
\textbf{Frequencies of most common AuNP synthesis precursors}. The most frequently extracted precursors using materials entity recognition (MER, ``Synthesis Recipe Extraction'') were inspected and compiled into a regular expression-based synonym map, which is housed in the available code repository (see ``Code Availability''). The precursors are binned by their function in AuNP synthesis and their presence in the number of publications employing seed-mediated growth or seedless methods is distinguished. Citrates are considered in their own category since their function varies depending on the method used, for instance citrates are used as a reducing agent and a ligand in Turkevich or Frens reduction, but only as a ligand in other reduction methods. Only the precursors appearing in more than 50 articles are shown. Precursors were counted once per article within the seed-mediated and seedless growth categories for this analysis to avoid double counting precursors which may be mentioned in a purchasing paragraph and a synthesis paragraph, both of which can sometimes be classified as a synthesis paragraph by our binary gold nanoparticle synthesis paragraph classifier.
}
\label{fig:precs}
\end{figure}

First, we present an overview and statistical breakdown of the precursors used for gold nanoparticle synthesis across the literature in Figure \ref{fig:precs}. Some measures were first taken to standardize the information extracted by MER (see ``Synthesis Recipe Extraction'') from each of these paragraphs. This included manually normalizing all synonyms for a given precursor to a single precursor name, as well as investigating and mapping variances of their token representations (e.g. from text-scraping errors, similar unicode characters, typos, etc.) to a single precursor name. The map consists of appropriately curated regex strings capturing these variations for a given precursor. This mapping is provided as a JSON file in the associated GitHub repository for this dataset (see ``Code Availability''). The presence of each precursor among seed-mediated and seedless growths is also reflected in this breakdown.
The overwhelming presence of $\textrm{HAuCl}_4$ is expected since this is the most prevalent gold source for synthesizing nanoparticles, with $\textrm{AuCl}_3$ and $\textrm{NaAuCl}_4$ following. 20 synthesis paragraphs were inspected that did not show any of these gold sources extracted as precursors. From these, it appeared that 15 paragraphs contained incomplete synthesis descriptions in the text. These were most often brief statements regarding the method of synthesis (e.g. ``AuNPs were synthesized through the Turkevich method...'') followed by a description of their resultant size. Although the synthesis information for these paragraphs was incomplete, they still often included successfully extracted morphological information that we consider valuable for the purposes of this dataset. The remaining 5 paragraphs showed issues with materials entity parsing (see ``Synthesis Recipe Extraction'') due to unusual syntactic structure (here, the gold precursor would be extracted and classified as \emph{other\_material} as opposed to \emph{precursor}, according to the data structure in Table \ref{tab:tab3}).
To better organize the distribution of precursors, we binned each according to their function in a given synthesis. Citrates were given their own bin since they can be used as either a ligand or both a reducing agent and ligand (as in Turkevich \cite{Turkevich} or Frens \cite{Frens} reduction), and because it is currently difficult to extract the specific role of a precursor using our language processing tools.  Strong and weak reducing agents were binned together, where $\textrm{NaBH}_4$ is used as a strong reducing agent while the other three are considered weak. 
This breakdown also indicates which precursors are frequently used for seed-mediated growths, like CTAB and $\textrm{AgNO}_3$ for the growth of nanorods \cite{AuNRSeedMech}. The common precursors used for seedless growths are often based on attested reduction methods like Turkevich or Frens reduction in this dataset, which both incorporate citrate-based precursors \cite{AuNPMethods}. 
The lower frequency for several of the precursors is likely due to their relatively recent introduction into the field, such as PVP which was used first in gold nanoparticle synthesis in 2017 to limit growth of nanorods as a capping agent \cite{PVPRod}.
The low presence of water is likely due to the manner in which precursors are extracted for this dataset using MER, which can extract precursor entities like ``water'' and ``$\textrm{H}_2\textrm{O}$'', but cannot infer water as a precursor from descriptions of solutions like ``aqueous''. 

\begin{figure}[ht]
\centering
\includegraphics[width=\linewidth]{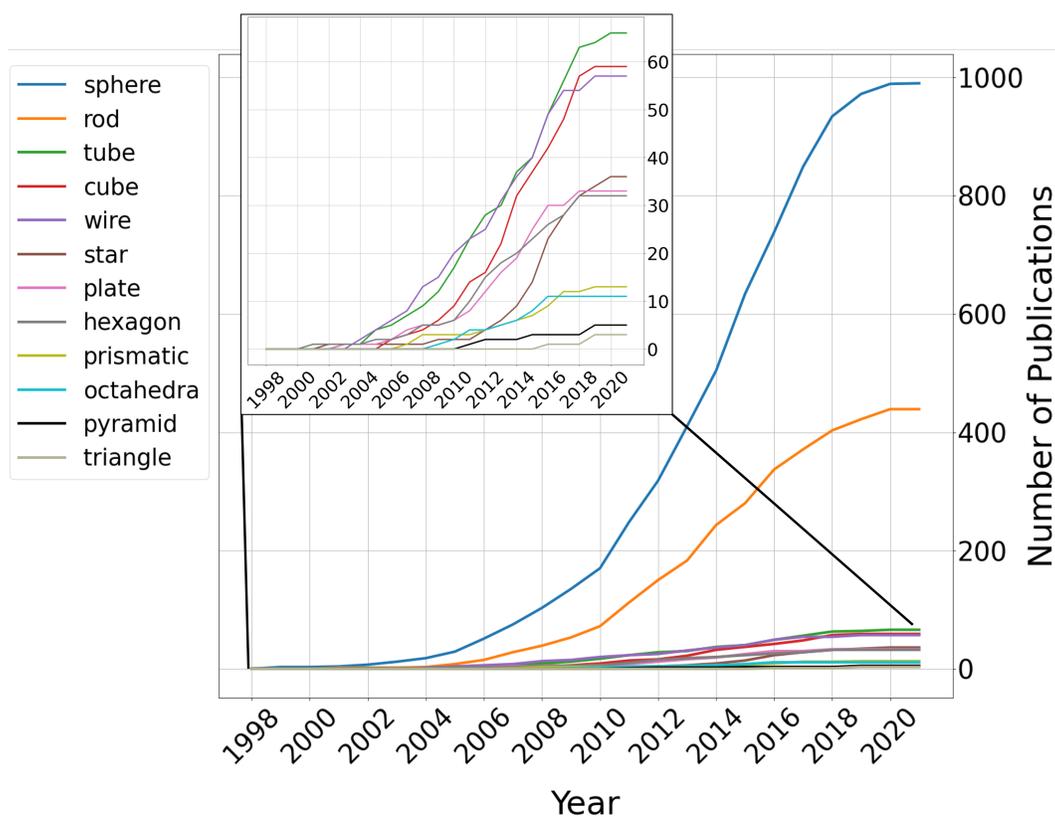}
\caption{
\textbf{Breakdown of reported AuNP morphologies discussed by year}. Similarly to the frequent precursors analysis, we compiled a regular expression-based synonym map for the most frequently extracted morphological entities and descriptors. The timeline represents the cumulative number of publications discussing each of the specified morphologies from 1998-2021. The entire plot represents 1,744 articles from the dataset, each of which discuss only one of the morphologies in the set depicted. 
}
\label{fig:morphs}
\end{figure}

Moving beyond synthesis details, we also analyze the breakdown of the morphologies discussed in the literature and how those have varied cumulatively across time. Figure \ref{fig:morphs} represents the proportion of the most discussed morphologies in the gold nanoparticle literature published between 1998 and 2021. For the purposes of this breakdown, only articles that discuss a single morphology are considered. Through this filtration, the breakdown consists of 1,744 articles out of the 5,154 in the dataset. Morphologies were determined using the \texttt{morphologies} and \texttt{descriptors} fields in the \texttt{morphological\_information} field, which combines multiple-entity strings from the MorphER extraction results. The synonym map used to normalize the extracted entities is available in the associated codebase, which was constructed in a similar manner to the precursors synonym map.
The strong presence of spherical particles across all years is due to their longevity in the field, being synthesized through a formal procedure by Faraday as early as 1857 \cite{Faraday}. Spherical particles are also straightforward to synthesize, with facile methods being pioneered by Turkevich \cite{Turkevich} and Frens \cite{Frens} in the 1950s and 1970s, respectively.
Rods are discussed in a quarter of collected publications, more than any of the other anisotropic shapes combined. This reflects the trends in the literature \cite{AuNRReview2}, mostly due to their highly tuneable optical properties and more recently developed convenient wet synthesis methods \cite{RodTricks}.

\begin{figure}[ht]
\centering
\includegraphics[width=\linewidth]{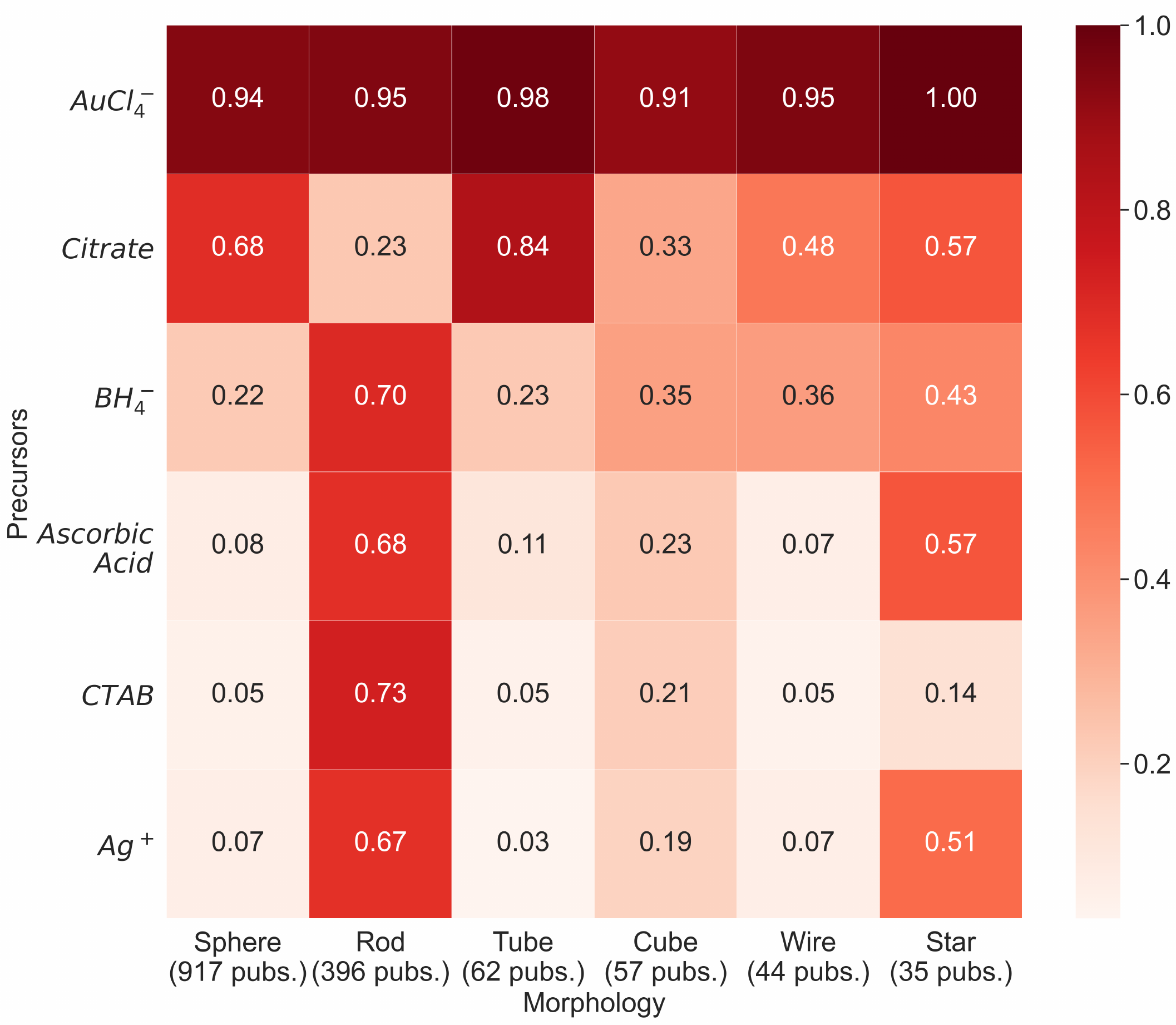}
\caption{
\textbf{Heatmap depicting correlation between precursors and resultant AuNP morphologies}. The heat illustrated in a given cell represents the fraction of morphologically-targeted articles (say, the fraction of sphere-related articles) which use that particular precursor among one or more precursors it uses in the recipe. For instance, the top left cell shows that more than 90\% of purely sphere-related AuNP synthesis papers use $\textrm{AuCl}_4^-$ as a precursor. "Citrate" also includes sodium citrate precursors. The entire heatmap describes 1,511 single morphology-targeted articles with at least one of the precursors or precursor ions shown on y-axis. 
}
\label{fig:precsmorphs}
\end{figure}

Finally, we explore correlations between the use of certain precursors and the target morphology of a given synthesis. Using the filtration process described above to consider only single morphology publications related to spheres, rods, tubes, cubes, wires, and stars yielded 1,647 publications. Assuming the one mentioned morphology is indeed the target, we developed a heat map presenting the proportion of select precursors and common precursor ions ($\textrm{AuCl}_4^-$, citrate, CTAB, $\textrm{BH}_4^-$, ascorbic acid, and $\textrm{Ag}^+$) mentioned in publications with each target morphology (Figure \ref{fig:precsmorphs}). In this plot, ``Citrate'' also contains sodium citrate precursors. The extracted precursors used in a given synthesis were matched against the precursor synonym bank discussed previously. With this additional filtration step, a total of 1,511 publications with only one of the select set of morphologies mentioned and also having at least one of the select precursors are shown in the heat map. A few general trends are reflected in this illustration. First, the frequent mentions of CTAB, $\textrm{Ag}^+$, ascorbic acid, and $\textrm{BH}_4^-$ are distinct for nanorod synthesis publications. In particular, the use of $\textrm{AgNO}_3$ and CTAB to control the quality and characteristics of gold nanorod growth is well-known to the nanoparticle synthesis field \cite{AuNRSeedMech}. $\textrm{AgNO}_3$ is used to control the aspect ratios of the rods and there was a recent shift in seed-mediated growth from citrate-capped gold seeds to CTAB-capped gold seeds because the latter showed an improvement on earlier particle formation limitations (e.g. noncylindrical rods, spherical impurities, etc.). Second, citrate is most prominantly used in the synthesis of spherical particles. As was discussed regarding the precursor breakdown earlier (Figure \ref{fig:precs}), citrate was used in the seminal experimental works by Turkevich \cite{Turkevich} and later by Frens \cite{Frens} as both a reducing and stabilizing agent. These methods are still among the most prominent for synthesizing spherical gold nanoparticles, as is reflected by Figure \ref{fig:precsmorphs}.


As was discussed in ``Synthesis Recipe Extraction'', the language processing tools and methods used to create this dataset were adapted from tools previously developed for the extraction of solid-state \cite{KononovaSolidState} and solution-based \cite{SolutionDataset} recipes. The experimental methods used for nanoparticle synthesis are distinct from other materials synthesis methods. This is particularly so from solid-state methods, but even holds for other solution-based synthesis methods. Because of this, we built additional extraction methods (see ``AuNP Synthesis Paragraph Classification'', ``Characterization Paragraph Classification'', and ``Characterization Entity Recognition (MorphER)''), on top of those that were used for the construction of text-mined solid-state and solution-based recipe datasets, to better handle such synthesis details. However, there are still some pitfalls in this combination of extraction methods that we are addressing for future iterations of this dataset. 
First, the order of synthesis actions is particularly important in seed-mediated nanoparticle synthesis, which is found to represent a substantial fraction of the major synthesis methods found in the literature (see Figure \ref{fig:precs}). This method is comprised of a seed solution preparation step, a growth solution preparation step, and a step that combines these two. Currently, our synthesis action extraction method cannot distinguish these three steps as separate synthesis procedures. Therefore, isolating specific synthesis procedures for the components of a given seed-mediated synthesis is difficult. Because the seed and growth solutions are often described as ingredients in the text, they can be captured in the \texttt{subject} field of the \texttt{procedure\_graph} (Table \ref{tab:tab4}), which is determined through dependency tree parsing. To address this issue, noun-phrases parsed in the \texttt{subject} field can be used to define the relevant synthesis constituent being manipulated or prepared, and thus separate the synthesis procedures into components for seed-mediated growth. 
Second, the current materials entity recognition model does not detect entities that do not contain specific material formulae or chemical names. Thus, neither ``AuNP seed solution'' nor ``growth solution'' would be detected in the sentence ``...3 mL of AuNP seed solution was mixed with 5 mL of growth solution to produce the final nanorods.'' Because of this, the corresponding amounts for each component of the synthesis cannot be extracted. Such information is important for seed-mediated growth, so we plan to address this by using the results of the aforementioned \texttt{subject} field and use seed and growth solution-related noun phrases as anchors for an additional material amounts extraction step if the paragraph describes seed-mediated synthesis.
Finally, there is currently no way to distinguish extracted morphologies as either the desired target morphology or just a morphology mentioned off-hand by the author. To address this, we plan to develop an additional layer on top of the current morphology entity recognition model that classifies those entities predicted to be \textbf{MOR} into either target (\textbf{TGT}) or miscellaneous morphologies (\textbf{MIS}), similar to the strategy used for materials entity recognition (see ``Synthesis Recipe Extraction''). 

\section*{Usage Notes}

The present dataset is provided as a single JSON file that can be read using all major programming languages (e.g. Python, Matlab, R, etc.). It is publicly available at \url{https://doi.org/10.6084/m9.figshare.16614262.v3} \cite{AuNP_Dataset}. No dependencies are required to access the contents of the dataset.

We invite users to utilize this dataset, among other applications, for the purposes of gold nanoparticle synthesis literature reviews or to query specific recipe protocols that achieve a desired morphology or size. 

This data descriptor defines a static version of the gold nanoparticle synthesis and characterization dataset; however, we intend to update the dataset in the repository below on a regular basis here: \url{https://github.com/CederGroupHub/text-mined-aunp-synthesis_public}. This will soon include updates addressing the issues discussed at the end of ``Dataset Mining'' as well as for morphological entity linking and linking target morphologies to specific recipe protocols. 

\section*{Code Availability}
\label{section: code_avail}

Scripts developed for the generation of this dataset as well as notebooks for example data analysis are available at \url{https://github.com/CederGroupHub/text-mined-aunp-synthesis_public}, along with an acknowledgement for this paper. The libraries use for this project are: \emph{ChemDataExtractor}, \emph{SpaCy}, \emph{scikit-learn}, \emph{gensim}, \emph{Tensorflow}, \emph{Keras}, \emph{PyTorch}, and \emph{Simple Transformers}.

\section*{Acknowledgements}

This work was funded by the U.S. Department of Energy, Office of Science, Office of Basic Energy Sciences, Materials Sciences and Engineering Division under Contract No. DE-AC02-05-CH11231 (D2S2 program KCD2S2). We thank Anna Sackmann, Rachael Samberg, and Timothy Vollmer (Science Data and Engineering Librarians at UC Berkeley) for assistance in obtaining Text and Data Mining agreements with the relevant publishers. We would also like to thank Sam Gleason, Xingzhi Wang, Jakob Dahl, Caitlin McCandler, John Dagdelen, Nicholas Walker, and Akshay Subramanian for valuable advice and discussion regarding the development of this pipeline and analysis of the data.

\section*{Author Contributions}

K.C. developed the extraction pipeline, annotated paragraphs for and trained the relevant paragraph classifiers and MorphER model, analyzed the data, and wrote the manuscript. A.T. provided guidance for approaches in developing the pipeline and developed framework for MorphER training. S.L. performed manual data inspection, advised on updates to extraction methods, and analyzed the data. Z.W. developed the material quantities extraction method. H.H. developed the Apache Solr search engine and trained MatBERT. T.H. developed materials entity recognition. O.K. developed synthesis action and conditions extraction method. A.J. supervised the project and wrote the manuscript. G.C. developed the approach, supervised the project, and wrote the manuscript. All authors contributed to the final manuscript. 

\section*{Competing Interests}

The authors declare no competing interests.

\bibliographystyle{naturemag}
\bibliography{refs}

\begin{thebibliography}{10}
\expandafter\ifx\csname url\endcsname\relax
  \def\url#1{\texttt{#1}}\fi
\expandafter\ifx\csname urlprefix\endcsname\relax\def\urlprefix{URL }\fi
\providecommand{\bibinfo}[2]{#2}
\providecommand{\eprint}[2][]{\url{#2}}

\bibitem{CancerDetect}
\bibinfo{author}{Liu, X.} \emph{et~al.}
\newblock \bibinfo{title}{A one-step homogeneous immunoassay for cancer
  biomarker detection using gold nanoparticle probes coupled with dynamic light
  scattering}.
\newblock \emph{\bibinfo{journal}{J. Am. Chem. Soc.}}
  \textbf{\bibinfo{volume}{130}}, \bibinfo{pages}{2780--2782}
  (\bibinfo{year}{2008}).

\bibitem{Semicond}
\bibinfo{author}{Dawson, A.} \& \bibinfo{author}{Kamat, P.~V.}
\newblock \bibinfo{title}{Semiconductor−metal nanocomposites. photoinduced
  fusion and photocatalysis of gold-capped {T}i{$\textrm{O}_2$}
  ({T}i{$\textrm{O}_2$}/gold) nanoparticles}.
\newblock \emph{\bibinfo{journal}{J. Phys. Chem. B}}
  \textbf{\bibinfo{volume}{105}}, \bibinfo{pages}{960--966}
  (\bibinfo{year}{2001}).

\bibitem{Cosmetics}
\bibinfo{author}{Kaul, S.}, \bibinfo{author}{Gulati, N.},
  \bibinfo{author}{Verma, D.}, \bibinfo{author}{Mukherjee, S.} \&
  \bibinfo{author}{Nagaich, U.}
\newblock \bibinfo{title}{Role of nanotechnology in cosmeceuticals: A review of
  recent advances}.
\newblock \emph{\bibinfo{journal}{Journal of Pharmaceutics}}
  \textbf{\bibinfo{volume}{2018}} (\bibinfo{year}{2018}).

\bibitem{PVPRod}
\bibinfo{author}{Requejo, K.~I.}, \bibinfo{author}{Liopo, A.~V.},
  \bibinfo{author}{Derry, P.~J.} \& \bibinfo{author}{Zubarev, E.~R.}
\newblock \bibinfo{title}{Accelerating gold nanorod synthesis with nanomolar
  concentrations of poly(vinylpyrrolidone)}.
\newblock \emph{\bibinfo{journal}{Langmuir}} \textbf{\bibinfo{volume}{33}},
  \bibinfo{pages}{12681--12688} (\bibinfo{year}{2017}).

\bibitem{AuNPSynthReview}
\bibinfo{author}{De~Souza, C.~D.}, \bibinfo{author}{Nogueira, B.~R.} \&
  \bibinfo{author}{Rostelato, M. E.~C.}
\newblock \bibinfo{title}{Review of the methodologies used in the synthesis
  gold nanoparticles by chemical reduction}.
\newblock \emph{\bibinfo{journal}{J. Alloys Compd.}}
  \textbf{\bibinfo{volume}{789}}, \bibinfo{pages}{714--740}
  (\bibinfo{year}{2019}).

\bibitem{AuNPShapeControl}
\bibinfo{author}{Grzelczak, M.}, \bibinfo{author}{P\'erez-Juste, J.},
  \bibinfo{author}{Mulvaney, P.} \& \bibinfo{author}{Liz-Marz\'an, L.~M.}
\newblock \bibinfo{title}{Shape control in gold nanoparticle synthesis}.
\newblock \emph{\bibinfo{journal}{Chem. Soc. Rev.}}
  \textbf{\bibinfo{volume}{37}}, \bibinfo{pages}{1783--1791}
  (\bibinfo{year}{2008}).

\bibitem{AuNPShapeControl2}
\bibinfo{author}{Personick, M.~L.} \& \bibinfo{author}{Mirkin, C.~A.}
\newblock \bibinfo{title}{Making sense of the mayhem behind shape control in
  the synthesis of gold nanoparticles}.
\newblock \emph{\bibinfo{journal}{J. Am. Chem. Soc.}}
  \textbf{\bibinfo{volume}{135}}, \bibinfo{pages}{18238--18247}
  (\bibinfo{year}{2013}).

\bibitem{FormationModel}
\bibinfo{author}{Agunloye, E.}, \bibinfo{author}{Panariello, L.},
  \bibinfo{author}{Gavriilidis, A.} \& \bibinfo{author}{Mazzei, L.}
\newblock \bibinfo{title}{A model for the formation of gold nanoparticles in
  the citrate synthesis method}.
\newblock \emph{\bibinfo{journal}{Chem. Eng. Sci.}}
  \textbf{\bibinfo{volume}{191}}, \bibinfo{pages}{318--331}
  (\bibinfo{year}{2018}).

\bibitem{AuNRReview2}
\bibinfo{author}{Lohse, S.~E.} \& \bibinfo{author}{Murphy, C.~J.}
\newblock \bibinfo{title}{The quest for shape control: A history of gold
  nanorod synthesis}.
\newblock \emph{\bibinfo{journal}{Chem. Mater.}} \textbf{\bibinfo{volume}{25}},
  \bibinfo{pages}{1250--1261} (\bibinfo{year}{2013}).

\bibitem{DFTAu12}
\bibinfo{author}{Mukhamedzyanova, D.~F.}, \bibinfo{author}{Ratmanova, N.~K.},
  \bibinfo{author}{Pichugina, D.~A.} \& \bibinfo{author}{Kuz'menko, N.~E.}
\newblock \bibinfo{title}{A structural and stability evaluation of
  {$\textrm{Au}_{12}$}}.
\newblock \emph{\bibinfo{journal}{J. Phys. Chem. C}}
  \textbf{\bibinfo{volume}{116}}, \bibinfo{pages}{11507--11518}
  (\bibinfo{year}{2012}).

\bibitem{DFTLigand}
\bibinfo{author}{Domingo, M.}, \bibinfo{author}{Shahrokhi, M.},
  \bibinfo{author}{Remediakis, I.} \& \bibinfo{author}{Lopez, N.}
\newblock \bibinfo{title}{Shape control in gold nanoparticles by n-containing
  ligands: Insights from density functional theory and wulff constructions}.
\newblock \emph{\bibinfo{journal}{Top. Catal.}} \textbf{\bibinfo{volume}{61}},
  \bibinfo{pages}{412--418} (\bibinfo{year}{2018}).

\bibitem{ProtMole}
\bibinfo{author}{Chakraborty, I.} \& \bibinfo{author}{Pradeep, T.}
\newblock \bibinfo{title}{Atomically precise clusters of noble metals: Emerging
  link between atoms and nanoparticles}.
\newblock \emph{\bibinfo{journal}{Chem. Rev.}} \textbf{\bibinfo{volume}{117}},
  \bibinfo{pages}{8208--8271} (\bibinfo{year}{2017}).

\bibitem{Continuum}
\bibinfo{author}{Talapin, D.~V.}, \bibinfo{author}{Rogach, A.~L.},
  \bibinfo{author}{Haase, M.} \& \bibinfo{author}{Weller, H.}
\newblock \bibinfo{title}{Evolution of an ensemble of nanoparticles in a
  colloidal solution: Theoretical study}.
\newblock \emph{\bibinfo{journal}{J. Phys. Chem. B}}
  \textbf{\bibinfo{volume}{105}}, \bibinfo{pages}{12278--12285}
  (\bibinfo{year}{2001}).

\bibitem{MatDiscovery}
\bibinfo{author}{Ren, F.} \emph{et~al.}
\newblock \bibinfo{title}{Accelerated discovery of metallic glasses through
  iteration of machine learning and high-throughput experiments}.
\newblock \emph{\bibinfo{journal}{Sci. Adv.}} \textbf{\bibinfo{volume}{4}},
  \bibinfo{pages}{4} (\bibinfo{year}{2018}).

\bibitem{FisherCrystalStructure}
\bibinfo{author}{Fischer, C.~C.}, \bibinfo{author}{Tibbetts, K.~J.},
  \bibinfo{author}{Morgan, D.} \& \bibinfo{author}{Ceder, G.}
\newblock \bibinfo{title}{Predicting crystal structure by merging data mining
  with quantum mechanics}.
\newblock \emph{\bibinfo{journal}{Nat. Mat.}} \textbf{\bibinfo{volume}{5}},
  \bibinfo{pages}{641--646} (\bibinfo{year}{2006}).

\bibitem{WestonNER}
\bibinfo{author}{Weston, L.} \emph{et~al.}
\newblock \bibinfo{title}{Named entity recognition and normalization applied to
  large-scale information extraction from the materials science literature}.
\newblock \emph{\bibinfo{journal}{J. Chem. Inf. Model.}}
  \textbf{\bibinfo{volume}{59}}, \bibinfo{pages}{3692--3702}
  (\bibinfo{year}{2019}).

\bibitem{AutoDetect}
\bibinfo{author}{Wang, X.} \emph{et~al.}
\newblock \bibinfo{title}{{AutoDetect-mNP}: An unsupervised machine learning
  algorithm for automated analysis of transmission electron microscope images
  of metal nanoparticles}.
\newblock \emph{\bibinfo{journal}{JACS Au}} \textbf{\bibinfo{volume}{1}},
  \bibinfo{pages}{316--327} (\bibinfo{year}{2021}).

\bibitem{SzymanskiXRD}
\bibinfo{author}{Szymanski, N.~J.}, \bibinfo{author}{Bartel, C.~J.},
  \bibinfo{author}{Zeng, Y.}, \bibinfo{author}{Tu, Q.} \&
  \bibinfo{author}{Ceder, G.}
\newblock \bibinfo{title}{Probabilistic deep learning approach to automate the
  interpretation of multi-phase diffraction spectra}.
\newblock \emph{\bibinfo{journal}{Chem. Mat.}} \textbf{\bibinfo{volume}{33}},
  \bibinfo{pages}{4204--4215} (\bibinfo{year}{2021}).

\bibitem{NPStructureDb}
\bibinfo{author}{Yan, X.}, \bibinfo{author}{Sedykh, A.}, \bibinfo{author}{Wang,
  W.}, \bibinfo{author}{Yan, B.} \& \bibinfo{author}{Zhu, H.}
\newblock \bibinfo{title}{Construction of a web-based nanomaterial database by
  big data curation and modeling friendly nanostructure annotations}.
\newblock \emph{\bibinfo{journal}{Nat. Comm.}} \textbf{\bibinfo{volume}{11}}
  (\bibinfo{year}{2020}).

\bibitem{LLNLNano}
\bibinfo{author}{Hiszpanski, A.~M.} \emph{et~al.}
\newblock \bibinfo{title}{Nanomaterial synthesis insights from machine learning
  of scientific articles by extracting, structuring, and visualizing
  knowledge}.
\newblock \emph{\bibinfo{journal}{J. Chem. Inf. Model.}}
  \textbf{\bibinfo{volume}{6}}, \bibinfo{pages}{2876--2887}
  (\bibinfo{year}{2020}).

\bibitem{KrallingerIR2017}
\bibinfo{author}{Krallinger, M.}, \bibinfo{author}{Rabal, O.},
  \bibinfo{author}{Lourenço, A.}, \bibinfo{author}{Oyarzabal, J.} \&
  \bibinfo{author}{Valencia, A.}
\newblock \bibinfo{title}{Information retrieval and text mining technologies
  for chemistry}.
\newblock \emph{\bibinfo{journal}{Chem. Rev.}} \textbf{\bibinfo{volume}{117}},
  \bibinfo{pages}{7673--7761} (\bibinfo{year}{2017}).

\bibitem{ChemTagger}
\bibinfo{author}{Hawizy, L.}, \bibinfo{author}{Jessop, D.~M.},
  \bibinfo{author}{Adams, N.} \& \bibinfo{author}{Murray-Rust, P.}
\newblock \bibinfo{title}{{ChemicalTagger}: A tool for semantic text-mining in
  chemistry}.
\newblock \emph{\bibinfo{journal}{J. Cheminformatics}}
  \textbf{\bibinfo{volume}{3}}, \bibinfo{pages}{17} (\bibinfo{year}{2011}).

\bibitem{CDE}
\bibinfo{author}{Swain, M.~C.} \& \bibinfo{author}{Cole, J.~M.}
\newblock \bibinfo{title}{{ChemDataExtractor}: A toolkit for automated
  extraction of chemical information from the scientific literature}.
\newblock \emph{\bibinfo{journal}{J. Chem. Inf. Model.}}
  \textbf{\bibinfo{volume}{56}}, \bibinfo{pages}{1894--1904}
  (\bibinfo{year}{2016}).

\bibitem{KononovaTMReview}
\bibinfo{author}{Kononova, O.} \emph{et~al.}
\newblock \bibinfo{title}{Opportunities and challenges of text mining in
  materials research}.
\newblock \emph{\bibinfo{journal}{iScience}} \textbf{\bibinfo{volume}{24}},
  \bibinfo{pages}{3} (\bibinfo{year}{2021}).

\bibitem{OlivettiReview}
\bibinfo{author}{Olivetti, E.} \emph{et~al.}
\newblock \bibinfo{title}{Data-driven materials research enabled by natural
  language processing}.
\newblock \emph{\bibinfo{journal}{Appl. Phys. Rev.}}
  \textbf{\bibinfo{volume}{7}}, \bibinfo{pages}{041317} (\bibinfo{year}{2020}).

\bibitem{KimInsights}
\bibinfo{author}{Kim, E.} \emph{et~al.}
\newblock \bibinfo{title}{Materials synthesis insights from scientific
  literature via text extraction and machine learning}.
\newblock \emph{\bibinfo{journal}{Chem. Mater}} \textbf{\bibinfo{volume}{29}},
  \bibinfo{pages}{9436--9444} (\bibinfo{year}{2017}).

\bibitem{Tshitoyan2019}
\bibinfo{author}{Tshitoyan, V.} \emph{et~al.}
\newblock \bibinfo{title}{Unsupervised word embeddings capture latent knowledge
  from materials science literature}.
\newblock \emph{\bibinfo{journal}{Nature}} \textbf{\bibinfo{volume}{571}},
  \bibinfo{pages}{95--98} (\bibinfo{year}{2019}).

\bibitem{KononovaSolidState}
\bibinfo{author}{Kononova, O.} \emph{et~al.}
\newblock \bibinfo{title}{Text-mined dataset of inorganic materials synthesis
  recipes}.
\newblock \emph{\bibinfo{journal}{Sci. Data}} \textbf{\bibinfo{volume}{6}},
  \bibinfo{pages}{203} (\bibinfo{year}{2019}).

\bibitem{jurafskynlp}
\bibinfo{author}{Jurafsky, D.} \& \bibinfo{author}{Martin, J.}
\newblock \emph{\bibinfo{title}{Speech and Language Processing: An Introduction
  to Natural Language Processing, Computational Linguistics, and Speech
  Recognition}}.
\newblock Prentice Hall Series in Artificial Intelligence
  (\bibinfo{publisher}{Pearson Prentice Hall}, \bibinfo{year}{2009}).

\bibitem{BERT}
\bibinfo{author}{Devlin, J.}, \bibinfo{author}{Chang, M.-W.},
  \bibinfo{author}{Lee, K.} \& \bibinfo{author}{Toutanova, K.}
\newblock \bibinfo{title}{{BERT}: Pre-training of deep bidirectional
  transformers for language understanding}.
\newblock In \emph{\bibinfo{booktitle}{Proceedings of the 2019 Conference of
  the North {A}merican Chapter of the Association for Computational
  Linguistics: Human Language Technologies, Volume 1 (Long and Short Papers)}},
  \bibinfo{pages}{4171--4186} (\bibinfo{publisher}{Association for
  Computational Linguistics}, \bibinfo{address}{Minneapolis, Minnesota},
  \bibinfo{year}{2019}).

\bibitem{MatBERTNER}
\bibinfo{author}{Walker, N.} \emph{et~al.}
\newblock \bibinfo{title}{The impact of domain-specific pre-training on named
  entity recognition tasks in materials science}.
\newblock \emph{\bibinfo{journal}{Patterns}}  (\bibinfo{year}{2022}).

\bibitem{LDA}
\bibinfo{author}{Blei, D.~M.}, \bibinfo{author}{Ng, A.~Y.} \&
  \bibinfo{author}{Jordan, M.~I.}
\newblock \bibinfo{title}{Latent dirichlet allocation}.
\newblock \emph{\bibinfo{journal}{J. of Mach. Learn. Res.}}
  \textbf{\bibinfo{volume}{3}}, \bibinfo{pages}{993--1022}
  (\bibinfo{year}{2003}).

\bibitem{HeCER}
\bibinfo{author}{He, T.} \emph{et~al.}
\newblock \bibinfo{title}{Similarity of precursors in solid-state synthesis as
  text-mined from scientific literature}.
\newblock \emph{\bibinfo{journal}{Chem. Mat.}} \textbf{\bibinfo{volume}{32}},
  \bibinfo{pages}{7861--7873} (\bibinfo{year}{2020}).

\bibitem{HuoParaClass}
\bibinfo{author}{Huo, H.} \emph{et~al.}
\newblock \bibinfo{title}{Semi-supervised machine-learning classification of
  materials synthesis procedures}.
\newblock \emph{\bibinfo{journal}{Npj Comput. Mater.}}
  \textbf{\bibinfo{volume}{5}}, \bibinfo{pages}{62} (\bibinfo{year}{2019}).

\bibitem{mikolov2013distributed}
\bibinfo{author}{Mikolov, T.}, \bibinfo{author}{Sutskever, I.},
  \bibinfo{author}{Chen, K.}, \bibinfo{author}{Corrado, G.} \&
  \bibinfo{author}{Dean, J.}
\newblock \bibinfo{title}{Distributed representations of words and phrases and
  their compositionality} (\bibinfo{year}{2013}).
\newblock \eprint{1310.4546}.

\bibitem{rehurek_lrec}
\bibinfo{author}{{\v R}eh{\r u}{\v r}ek, R.} \& \bibinfo{author}{Sojka, P.}
\newblock \bibinfo{title}{Software framework for topic modelling with large
  corpora}.
\newblock In \emph{\bibinfo{booktitle}{{Proceedings of the LREC 2010 Workshop
  on New Challenges for NLP Frameworks}}}, \bibinfo{pages}{45--50}
  (\bibinfo{publisher}{ELRA}, \bibinfo{address}{Valletta, Malta},
  \bibinfo{year}{2010}).

\bibitem{SpaCy}
\bibinfo{author}{Honnibal, M.} \& \bibinfo{author}{Johnson, M.}
\newblock \bibinfo{title}{An improved non-monotonic transition system for
  dependency parsing}.
\newblock In \emph{\bibinfo{booktitle}{Proceedings of the 2015 Conference on
  Empirical Methods in Natural Language Processing}},
  \bibinfo{pages}{1373--1378} (\bibinfo{publisher}{Association for
  Computational Linguistics}, \bibinfo{address}{Lisbon, Portugal},
  \bibinfo{year}{2015}).

\bibitem{NLTK}
\bibinfo{author}{Bird, E.~L., Steven} \& \bibinfo{author}{Klein, E.}
\newblock \emph{\bibinfo{title}{Natural Language Processing with Python.
  O'Reilly Media Inc}} (\bibinfo{year}{2009}).

\bibitem{AuNRReview}
\bibinfo{author}{Huang, X.}, \bibinfo{author}{Neretina, S.} \&
  \bibinfo{author}{El-Sayed, M.~A.}
\newblock \bibinfo{title}{Gold nanorods: From synthesis and properties to
  biological and biomedical applications}.
\newblock \emph{\bibinfo{journal}{Adv. Mat.}} \textbf{\bibinfo{volume}{21}},
  \bibinfo{pages}{4880--4910} (\bibinfo{year}{2009}).

\bibitem{AuNP_Dataset}
\bibinfo{author}{Cruse, K.} \emph{et~al.}
\newblock \bibinfo{title}{{Text-mined AuNP Synthesis Recipes Dataset}}.
\newblock
  \bibinfo{howpublished}{\url{https://doi.org/10.6084/m9.figshare.16614262.v3}}
  (\bibinfo{year}{2021}).
\newblock \bibinfo{note}{Dataset}.

\bibitem{SolutionDataset}
\bibinfo{author}{Wang, Z.} \emph{et~al.}
\newblock \bibinfo{title}{{Dataset of solution-based inorganic materials
  synthesis recipes extracted from the scientific literature}}.
\newblock \bibinfo{howpublished}{Accepted to \emph{Sci. Data}. Preprint at
  \url{https://doi.org/10.48550/arXiv.2111.10874}} (\bibinfo{year}{2022}).

\bibitem{Turkevich}
\bibinfo{author}{Turkevich, J.}, \bibinfo{author}{Stevenson, P.~C.} \&
  \bibinfo{author}{Hillier, J.}
\newblock \bibinfo{title}{A study of the nucleation and growth processes in the
  synthesis of colloidal gold}.
\newblock \emph{\bibinfo{journal}{Discuss. Faraday Soc.}}
  \textbf{\bibinfo{volume}{11}}, \bibinfo{pages}{55--75}
  (\bibinfo{year}{1951}).

\bibitem{Frens}
\bibinfo{author}{Frens, G.}
\newblock \bibinfo{title}{Controlled nucleation for the regulation of the
  particle size in monodisperse gold suspensions}.
\newblock \emph{\bibinfo{journal}{Nat. Phys. Sci.}}
  \textbf{\bibinfo{volume}{241}}, \bibinfo{pages}{20--22}
  (\bibinfo{year}{1973}).

\bibitem{AuNRSeedMech}
\bibinfo{author}{Nikoobakht, B.} \& \bibinfo{author}{El-Sayed, M.~A.}
\newblock \bibinfo{title}{Preparation and growth mechanism of gold nanorods
  ({NR}s) using seed-mediated growth method}.
\newblock \emph{\bibinfo{journal}{Chem. Mater.}} \textbf{\bibinfo{volume}{15}}
  (\bibinfo{year}{2003}).

\bibitem{AuNPMethods}
\bibinfo{author}{Herizchi, R.}, \bibinfo{author}{Abbasi, E.},
  \bibinfo{author}{Milani, M.} \& \bibinfo{author}{Akbarzadeh, A.}
\newblock \bibinfo{title}{Current methods for synthesis of gold nanoparticles}.
\newblock \emph{\bibinfo{journal}{Artificial Cells, Nanomedicine, and
  Biotechnology}} \textbf{\bibinfo{volume}{44}}, \bibinfo{pages}{596--602}
  (\bibinfo{year}{2016}).

\bibitem{Faraday}
\bibinfo{author}{Faraday, M.}
\newblock \bibinfo{title}{X. the bakerian lecture. - experimental relations of
  gold (and other metals) to light}  (\bibinfo{year}{1857}).

\bibitem{RodTricks}
\bibinfo{author}{Scarabelli, L.}, \bibinfo{author}{S\'anchez-Iglesias, A.},
  \bibinfo{author}{P\'erez-Juste, J.} \& \bibinfo{author}{Liz-Marzan, L.~M.}
\newblock \bibinfo{title}{A "tips and tricks" practical guide to the synthesis
  of gold nanorods}.
\newblock \emph{\bibinfo{journal}{J. Phys. Chem. Lett}}
  \textbf{\bibinfo{volume}{6}}, \bibinfo{pages}{4270--4279}
  (\bibinfo{year}{2015}).

\bibitem{ULSA}
\bibinfo{author}{Wang, Z.} \emph{et~al.}
\newblock \bibinfo{title}{{ULSA}: Unified language of synthesis actions for
  representation of synthesis protocols}.
\newblock \bibinfo{howpublished}{Preprint at
  \url{https://doi.org/10.48550/arXiv.2201.09329}} (\bibinfo{year}{2022}).

\end{thebibliography}

\linespread{1.0}
\section*{Tables}




\begin{table}[ht]
\centering
\begin{threeparttable}[b]
\begin{tabular}{|p{1.8in}|c|l|}
\hline
\textbf{Data description} & \textbf{Data Key Label} & \textbf{Data Type} \\
\hline
DOI of the original paper & \texttt{doi} & \emph{string} \\
\hline
List of constituent paragraphs and extracted data & \texttt{paragraphs} & \emph{list} of Objects (\emph{dict})\tnote{1} \\
\hline
Year of publication & \texttt{publication\_year} & \emph{int} \\
\hline
Number of citations & \texttt{times\_referenced} & \emph{int} \\
\hline
\end{tabular}
\begin{tablenotes}
 \item[1] Contents of paragraphs shown in Table \ref{tab:tab3}
\end{tablenotes}
\end{threeparttable}
\caption{
Format for highest article-level of each data record: description, key label, data type. 
}
\label{tab:tab2}
\end{table}


\begin{table}[ht]
\centering
\begin{threeparttable}[b]
\begin{tabular}{|p{1.9in}|c|l|}
\hline
\textbf{Data description} & \textbf{Data Key Label} & \textbf{Data Type} \\
\hline
Unique paragraph hash & \texttt{\_id} & \emph{string} \\
\hline
Whether paragraph contains characterization information & \texttt{contains\_characterization} & \emph{bool} \\
\hline
Whether paragraph contains synthesis recipe & \texttt{contains\_recipe} & \emph{bool} \\
\hline
\multirow{4}{1.6in}{Materials and quantities contained in paragraph\tnote{1}} & \texttt{materials\_and\_quantities} & \emph{list} of Objects (\emph{dict}): \\
 & & -\texttt{material}: \emph{string} \\
 & & -\texttt{amount}: \emph{list} of Objects:\\
 & & - -\texttt{value}: \emph{float} \\
 & & - -\texttt{unit}: \emph{string} \\
\hline
\multirow{4}{1.6in}{Condensed morphological entities in paragraph\tnote{2}}& \texttt{morphological\_information} & Object (\emph{dict}): \\
 & & - \texttt{descriptors}: \emph{list} \\
 & & - \texttt{measurements}: \emph{list} \\
 & & - \texttt{morphologies}: \emph{list} \\
 & & - \texttt{seeds}: \emph{list} \\
 & & - \texttt{sizes}: \emph{list} \\
 & & - \texttt{units}: \emph{list} \\
\hline
\multirow{6}{1.6in}{Morphological entities and token locations in paragraph\tnote{2}}& \texttt{morphology\_ner\_tokens} & \emph{list} of Objects (\emph{dict}): \\
 & & - \texttt{annotation}: \emph{string} \\
 & & - \texttt{start}: \emph{int} \\
 & & - \texttt{end}: \emph{int} \\
 & & - \texttt{text}: \emph{string} \\
 & & \\
\hline
Whether or not the AuNP synthesis is seed-mediated\tnote{1} & \texttt{seed\_mediated} & \emph{bool} \\
\hline
List of constituent sentences and extracted data\tnote{1} & \texttt{sentences} & \emph{list} of Objects (\emph{dict})\tnote{3} \\
\hline
\multirow{4}{1.6in}{Synthesis actions and conditions\tnote{1}} & \texttt{synth\_actions} & \emph{list} of Objects (\emph{dict}): \\
 & & -\texttt{conditions}: Object (\emph{dict}): \\
 & & - -\texttt{temperature}: Object(\emph{dict})\tnote{4} \\
 & & - -\texttt{time}: Object(\emph{dict})\tnote{4} \\
 & & -\texttt{string}: \emph{string} \\
 & & -\texttt{subject}: \emph{string} \\
 & & -\texttt{type}: \emph{string} \\
\hline
Snippet of paragraph text & \texttt{text} & \emph{string} \\
\hline
\end{tabular}
\begin{tablenotes}
 \item[1] Only if \texttt{contains\_recipe} is \texttt{true}
 \item[2] Only if \texttt{contains\_characterization} or \texttt{contains\_recipe} is \texttt{true}
 \item[3] Contents of paragraphs shown in Table \ref{tab:tab4}
 \item[4] \{\texttt{value}: \emph{list}, \texttt{unit}: \emph{string}, \texttt{max\_value}: \emph{float}, \texttt{min\_value}: \emph{float}\}
\end{tablenotes}
\end{threeparttable}
\caption{
Format for lower paragraph-level of each data record: description, key label, data type. 
}
\label{tab:tab3}
\end{table}


\begin{table}[ht]
\centering
\begin{threeparttable}[b]
\begin{tabular}{|p{1.8in}|c|l|}
\hline
\textbf{Data description} & \textbf{Data Key Label} & \textbf{Data Type} \\
\hline
All material entities & \texttt{all\_materials} & \emph{list} of Objects (\emph{dict})\tnote{1} \\
\hline
Non-precursors and non-target material entities & \texttt{other\_materials} & \emph{list}\\
\hline
Precursor material entities & \texttt{precursors} & \emph{list} of Objects (\emph{dict})\tnote{1} \\
\hline
\multirow{4}{1.6in}{Sequence of synthesis operations and conditions} & \texttt{procedure\_graph} & \emph{list} of Objects (\emph{dict}) \\
 & & - \texttt{env\_toks}: \emph{list} \\
 & & - \texttt{op\_token}: \emph{string} \\
 & & - \texttt{op\_type}: \emph{string} \\
 & & - \texttt{ref\_op}: \emph{bool} \\
 & & - \texttt{subject}: \emph{string} \\
 & & - \texttt{temp\_values}: \emph{list} of Objects (\emph{dict})\tnote{2} \\
 & & - \texttt{time\_values}: \emph{list} of Objects (\emph{dict})\tnote{2} \\
\hline
Target material entities & \texttt{target} & \emph{list} of Objects (\emph{dict})\tnote{1} \\
\hline
\end{tabular}
\begin{tablenotes}
 \item[1] \{\texttt{material}: \emph{string}, \texttt{amount}: [\{\texttt{value}: \emph{float}, \texttt{unit}: \emph{string}\}]\}
 \item[2] \{\texttt{max}: \emph{float}, \texttt{min}: \emph{float}, \texttt{tok\_ids}: \emph{list}, \texttt{units}: \emph{string}, \texttt{values}: \emph{list}\}
\end{tablenotes}
\end{threeparttable}
\caption{
Format for lowest sentence-level of each data record: description, key label, data type. 
}
\label{tab:tab4}
\end{table}


\begin{table}[ht]
\centering
\begin{threeparttable}[b]
\begin{tabular}{|p{0.3\linewidth}|p{0.28\linewidth}|l|}
\hline
\textbf{Pipeline Component} & \textbf{ML Method} & \textbf{F1: (precision|recall)} \\
\hline
Article filtering & regex/tf-idf & 0.96:  (1.00|0.92) \\
\hline
Synthesis paragraph classification & BERT classification & 0.90: (0.96|0.85)\\
\hline
Characterization paragraph classification & BERT classification & 0.90: (0.93|0.87)\\
\hline
\multirow{1}{1.6in}{Materials Entity Recognition} & BiLSTM+CRF (MatBERT embeddings) & \\
& & 0.95:(0.95|0.95) - materials\tnote{1} \\
& & 0.90:(0.89|0.91) - precursors\tnote{1}  \\
& & 0.85:(0.86|0.83) - targets\tnote{1} \\
\hline
\multirow{1}{1.6in}{Morphology Entity Recognition} & Fine-tuned MatBERT NER model & 0.87:(0.89|0.84) - Micro average\\
& & 0.92:(0.90|0.95) - MOR (morphology)\\
& & 0.56:(0.70|0.52) - DES (descriptor)  \\
& & 0.70:(0.83|0.64) - MES (measurement) \\
& & 0.69:(0.81|0.62) - SIZ (size value)  \\
& & 0.91:(0.94|0.91) - UNT (unit)  \\
\hline
Synthesis actions\tnote{2} & BiLSTM (Word2Vec embeddings) & 0.89 (0.90|0.88)\\
\hline
\multirow{1}{1.6in}{Synthesis conditions\tnote{3}} & Rule-based & \\
- Temperature & & 0.94: (0.97|0.92) \\
- Time & & 0.93: (0.98|0.89) \\
\hline
Material quantities\tnote{3} & Rule-based & 0.87: (0.90|0.85) \\
\hline
Seed-mediated tag & Rule-based & 1.00: (1.00|1.00) \\
\hline
\end{tabular}
\begin{tablenotes}
 \item[1] Metrics from He et al. \cite{HeCER}
 \item[2] Metrics from associated manuscript on synthesis actions extraction \cite{ULSA}
 \item[3] Metrics from accepted publication on solution synthesis extraction \cite{SolutionDataset}
\end{tablenotes}
\end{threeparttable}
\caption{
Text extraction model accuracies
}
\label{tab:tab5}
\end{table}
\section*{Figures}
\linespread{1.5}

\end{document}